\documentclass[11pt]{article}

\usepackage{amssymb}
\usepackage{amsmath}
\usepackage{graphicx}
\usepackage{color}
\usepackage{multirow}
\usepackage{fullpage}

\newcommand{\cA}{{\cal A}}
\newcommand{\cO}{{\cal O}}

\newcommand{\cC}{{\cal C}}

\newcommand{\cG}{{\cal G}}

\newcommand{\cV}{{\cal V}}

\newcommand{\qed}{\hfill $\square$ \smallbreak}

\newenvironment{proofof}{\noindent{\bf Proof of}}{\qed}

\begin{document}

\title{{\bf Deterministic Rendezvous Algorithms }}

\author{
Andrzej Pelc \footnotemark[1]
}
\date{ }
\maketitle
\def\thefootnote{\fnsymbol{footnote}}

\footnotetext[1]{ \noindent
D\'epartement d'informatique, Universit\'e du Qu\'ebec en Outaouais, Gatineau,
Qu\'ebec J8X 3X7, Canada.
E-mail:  {\tt pelc@uqo.ca}\\
Partially supported by NSERC discovery grant 8136 -- 2013 and
by the Research Chair in Distributed Computing at the
Universit\'e du Qu\'{e}bec en Outaouais.
}

\begin{abstract}

The task of  rendezvous (also called {\em gathering}) calls for a meeting of two or more mobile entities, starting from different positions in some environment.
Those entities are called mobile agents or robots, and the  environment can be a network modeled as a graph or a terrain in the plane,
possibly with obstacles. The rendezvous problem has been studied in many different scenarios. Two among many adopted assumptions particularly
influence the methodology to be used to accomplish rendezvous. One of the assumptions specifies whether the agents in their navigation can see
something apart from parts of the environment itself, for example other agents or marks left by them. The other assumption concerns the way in which the entities move:
it can be either deterministic or randomized. In this paper we survey results on deterministic rendezvous of agents that cannot see the other agents prior
to meeting them, and cannot leave any marks.

\vspace*{0.5cm}

\noindent
{\bf keywords:} mobile agent, rendezvous, deterministic, network, graph, terrain, plane

\vspace*{7cm}

\end{abstract}

\thispagestyle{empty}

\pagebreak

\section{Introduction}

How to meet in an unknown environment? This question has to be answered in many applications. The most obvious and commonly encountered are those
where the entities that have to meet are part of the natural world: they are humans or animals. One of the examples cited in  \cite{alpern02b} is the
Astronaut Problem, in which two astronauts land in distant places
on a planet, without any orientation, and have to  minimize the expected time of getting together. More common examples of situations when humans have to meet is the
task of finding a lost hiker in the mountains by rescuers, or meeting guests at the airport by their host. 
 Schelling  \cite{schelling60} studied issues related to rendezvous problems: two players want to meet
in an unknown town and have only one attempt to make. Schelling emphasized the need of finding  ``focal points'' (such as the main station, or the central square
of the town) that are likely to be chosen by both players (without previous agreement), due to their common cultural background. However, in algorithmic rendezvous problems, focal points often do not exist, when agents have to meet in the empty plane or in a highly symmetric network.  Rendezvous tasks are also frequent in the animal world, such as
gathering of migratory birds or undersea animals, or  penguin parents finding their offspring when they come back with food.

In computer science applications, the most interesting cases concern human-made agents. The first example of such mobile agents are
autonomous mobile robots that start in different locations of a planar terrain or a labyrinth, and have to meet. The reason of meeting can be to exchange samples of the ground previously collected by the robots, or exchange information obtained when exploring different parts of the terrain. The second example is that of software agents, i.e., mobile pieces of software that navigate in a computer
network in order to perform maintenance of its components or to collect data distributed in nodes of the network. Periodic meeting of software agents is necessary
to exchange collected data and plan further actions, possibly depending on those data.

Since rendezvous algorithms do not depend on the physical nature of the mobile entities executing them, but only on their perception capabilities, memory size, mobility characteristics and on the structure of the environment,  we will not distinguish between natural and artificial agents, and among the latter between mobile robots and software agents, and we will use the generic name of {\em agents}
regardless of whether the algorithm is to be applied to people, animals, mobile robots, or software agents.
In the case of more than two agents, the rendezvous problem is sometimes called {\em gathering}. For the sake of uniformity, we will call it {\em rendezvous} also in this
case, using the term {\em gathering} as a synonym.

Since rendezvous problems usually have to be solved without the help of any central monitor coordinating the actions of agents, these problems belong naturally to the 
domain of distributed computing. There are, however, many scenarios and models under which rendezvous has been studied. Two among many adopted assumptions particularly influence the methodology to be used to accomplish rendezvous. 

One of the assumptions specifies whether the agents in their navigation can see
something apart from the underlying environment itself, for example other agents or marks left by them.
Needless to say, such a capability significantly facilitates the task of rendezvous: for example, two agents seeing each other in the plane may meet approaching each other along the line joining them. The other assumption concerns the way in which the entities move:
it can be either deterministic or randomized. In a deterministic scenario, the initial positions of the agents are chosen by an adversary
which models the worst-case situation, and  each move of the agent is determined only by its current history that may include the identity of the agent (if any), and the part of the environment that the agent has seen to date. By contrast, in a randomized scenario, initial positions of the agents are often chosen at random and their moves may also involve coin tosses. Randomized rendezvous algorithms in networks often use random walks in the underlying graph.
The cost of rendezvous is also different in both scenarios: while in deterministic rendezvous the worst-case cost is usually considered (cost being defined
as the time or the length of the agents'  trajectories until rendezvous), in the randomized scenario it is the expected value of this quantity. In both cases the problem
is often to minimize the worst case (resp. expected) cost. Deterministic rendezvous problems usually require combinatorial tools, while randomized rendezvous often calls for analytic methods.

In this chapter we consider the task of rendezvous under the weaker variant of each of the above two assumptions. First we assume that the navigating agents do not
see other agents prior to their meeting and cannot leave any marks, and second, we consider only deterministic rendezvous algorithms. The decision of carving out this particular subdomain of the domain of rendezvous algorithms has two reasons. The first and the most obvious of them is the space limitation: the realm of rendezvous algorithms is very large and
covering all of it would require a large book rather than a chapter. The second reason is to avoid duplication of information contained in previous surveys and in other chapters of this book.

There are six main previous surveys concerning rendezvous. Chronologically the first of them is \cite{Al}, almost entirely contained in the second part of the excellent book \cite{alpern02b}. Both   \cite{Al} and  \cite{alpern02b}
concern randomized rendezvous, viewed from the operations research point of view. The third  survey is \cite{KKR}.
While its scope is large, the authors concentrate mainly on presenting rendezvous models and compare their underlying assumptions. 
The book \cite{KKM-book} deals mostly with rendezvous problems on the ring, only briefly mentioning other network topologies in this context. The survey \cite{Pe}
is the closest to the present chapter but it covers only deterministic rendezvous in networks. Finally, Chapter XXX of this book covers the rich domain of gathering 
agents in the plane, under a scenario where agents have contrasting capabilities: they have no memory but enjoy very strong perception capabilities -- they can 
periodically make snapshots in which they see other agents. 

The existence of the book \cite{alpern02b} is the main reason of our restriction to deterministic rendezvous algorithms, and chapter XXX of this book is the main reason of our restriction to rendezvous under the scenario where navigating agents cannot see much.

The present chapter differs from  \cite{Al} and  \cite{alpern02b} by concentrating on deterministic rather than on randomized scenarios;
it differs from \cite{KKR} by the level of details in treating the rendezvous problem:
besides presenting various models under which deterministic rendezvous is studied, we want to report precisely the results obtained under each of them, discussing how varying assumptions influence feasibility and complexity of rendezvous under various settings.  This chapter differs from  \cite{KKM-book} by discussing many different topologies, mostly arbitrary, even unknown graphs, rather than concentrating on a particular type of networks. As mentioned above, this chapter is the closest to
\cite{Pe}: the identity of the author may be a reason. What are the differences with respect to \cite{Pe}? The survey \cite{Pe} was published in 2012, reporting results until 2011. During these 7 years many new results concerning deterministic rendezvous appeared in the literature. Since many of them concerned the scenario where agents cannot leave any marks, and the scenario
allowing marking of the nodes was considered in \cite{Pe}, we decided to exclude this scenario from the present chapter, and add instead issues concerning rendezvous in the plane, as they often use rendezvous in a specific graph, the infinite grid, as a methodological tool.  There is some overlap of this chapter with \cite{Pe} but we tried to
discuss the older results covered there only in a cursory manner, concentrating on the new developments.

A remark is in order concerning the chronology of the reported results. In each case when a journal paper is available, we refer to it, as to the most definitive version. However, the journal version is sometimes published much later than the conference version in which a given result first appeared. This may scramble the precedence relations of results. In such cases we tried to mention the correct order of discoveries.

The rest of the chapter is organized as follows. In Section 2 we discuss various scenarios resulting from alternative assumptions adopted for the rendezvous problem, and mention methodological differences of the solutions in different models. The main dividing line in the entire body of research surveyed in this chapter is between
the two types of environment in which rendezvous has to take place: one type are networks modeled  as undirected graphs, the other is the plane or parts of it, with possible obstacles obstructing moves of the agents. Consequently, Section 3 covers rendezvous in networks and Section 4 is devoted to rendezvous in the plane.
Finally, Section 5 contains conclusions and open problems.

\section{Discussion of assumptions, models and methodology}

As announced in the introduction, we will consider the rendezvous problem in two different environments: in networks modeled as undirected graphs and in the plane or its parts. Among many alternative pairs of assumptions adopted for the study of rendezvous, this one is arguably the most basic, as it influences even the precise definition of rendezvous. In both cases, agents are modeled as points. In the first case they navigate in the graph, traversing its edges and visiting its nodes, and in the second case they move freely in the plane, possibly avoiding obstacles. In the case of rendezvous in networks, meeting of the agents is defined as being at the same node in the same time or as being in the same point of an edge at the same time. (We will further discuss submodels in which one or the other of these definitions is applied). In the case when agents move in the plane, they are also modeled as points. However, in particular where rendezvous has to take place in the empty plane, we cannot require
the agents to get to the same point of the plane at the same time. To see this, consider an easier problem, when one of the agents is inert and the other one has to find it.
(Rendezvous can always be reduced to this case, if the adversary decides to delay the start of one of the agents sufficiently long). Since the walking agent cannot see the inert agent prior to meeting, a correct algorithm would require to construct a trajectory, which is a curve in the plane passing through each given point after walking a finite distance. Such a curve does not exist. Hence, in the case of meeting in the plane, rendezvous is defined more loosely as {\em approach}: agents have to get at distance 1 of each other. An alternative, equivalent  way of modeling this situation is to assume that in the plane agents are not points but discs of diameter 1 and rendezvous is accomplished when these discs touch. Hence in the case of the plane we will concentrate on the task of approach. 

There is one exception to this change of definition. When agents are in a bounded part of the plane, possibly with ``holes'', i.e., impenetrable obstacles, then rendezvous
defined as getting to the same point at the same time can be required, because in this case agents can meet in a boundary point of the terrain or of one of the obstacles.
We will see such a situation in \cite{CKP3}.

We will now review the common assumptions used in the literature to consider rendezvous in each of the above scenarios. We first consider the network scenario.
The first common assumption is modeling the network as a simple undirected connected graph,
whose nodes represent processors, computers or stations of a communication network, or crossings of corridors of a labyrinth, depending on the application,
and links represent communication channels in a communication network, or corridors in a labyrinth. Modeling the network as an undirected graph captures the ability of the agents to move
in both directions along each link.  The assumption that the graph is simple (no self-loops or multiple edges) is motivated by most of the realistic applications, and connectivity
of the graph is a necessary condition on feasibility of rendezvous when starting from any initial positions: agents starting in different connected components could not  meet. Throughout the chapter, we use the term {\em graph} to mean a simple undirected connected graph.  

The second common assumption  is the anonymity of the underlying network: the absence of distinct names
of nodes that can be perceived by the navigating agents. There are two reasons for  seeking rendezvous algorithms that do not assume knowledge of node identities.
The first one is practical: while nodes may indeed have different labels, they may  refrain from informing the agents about them, e.g., for privacy or security reasons, or limited sensory capabilities of agents may prevent them from perceiving these names. The latter restriction is mostly applicable to mobile robots whose sensing device may be too weak to read such labels. The other reason for assuming anonymity of the network is methodological. If distinct names 
of nodes can be perceived by the agents, they can follow an algorithm which guides each of them to the node with the smallest  label and stop. Thus the rendezvous
problem becomes reducible to graph exploration, which has been well studied. 

The last common assumption concerns port numbers at each node. It is assumed that
a node of degree $d$ has ports $0,1,\dots, d-1$ corresponding to the incident edges. Ports at each node can be perceived by an agent visiting this node, but there is no coherence assumed between port labelings at different nodes. (In the case when such a coherence is assumed, for example in the case of an oriented grid, it will be explicitly mentioned). When an agent enters a node, it learns its degree and the port of entry. 
The reason for assuming the existence of port labelings accessible to agents is the following.
If an agent is unable to locally distinguish ports at a node, it may even be unable to
visit all neighbors of a node of degree at least 3. Indeed, after visiting the second neighbor, the agent cannot distinguish
the port leading to the first visited neighbor from the port leading to the unvisited one.  Thus an adversary may always force an agent to avoid all
but two edges incident to such a node.
Consequently, agents
initially located at two nodes of degree at least 3 might never be able to meet. From the practical point of view, assuming the existence of port numbers legible by the agents is a much less problematic assumption than assuming labels of nodes. First, the privacy and security reasons for not divulging node labels do not apply to port numbers. Second, the sensory capabilities of agents required to read port numbers are much smaller than those for reading node labels. For example, one port at a node can be marked by a ``red dot'' and then consecutive ports can have a pointer {\em next} from the preceding port. Reading this type of information requires minimal sensory capabilities. 

We will now review the common assumptions concerning rendezvous (i.e.,  approach) in the plane. These assumptions permit an agent to navigate in the plane in the absence of any visual information. It is assumed that the agent has a compass indicating North and that it has a measure of distance. These two features permit to
establish a system of orthogonal coordinates and permit to trace arbitrary angles. Consequently, the agent can execute instructions such as ``go at distance $x$
in direction $dir$'', where $dir$ is expressed as an angle from direction North. Notice that the compasses and the measures of length of different agents are not necessarily the same. If they are, this will be explicitly mentioned. 

We proceed to the overview of various alternative assumptions yielding different scenarios under which the rendezvous problem is usually considered, both in the network environment and in the plane.
There are two  such main pairs of assumptions. The first concerns the possibility to distinguish the agents: they can be either anonymous (i.e., identical), or each agent may have a distinct integer label  that it knows and can use as a parameter in the executed rendezvous algorithm which is common to all agents.
The second pair of alternative assumptions concerns time: agents may move either in a synchronous or in an asynchronous way. We will give the precise definitions later but, roughly speaking,
synchronous movement in graphs means that clocks of the agents tick at the same rate, one tick per round, and in each round an agent can either stay in the current node, or move to a neighbor. In the plane, synchronous movement means that the speed of agents is the same. In the asynchronous scenario, the speed of the agents may vary adversarially. (We will see that there is also a semi-synchronous
scenario, where speeds of agents are constant but possibly different). 

The above possible scenarios imply different methodological approach to rendezvous in each case.
The main problem that has to be solved in order to make deterministic rendezvous possible, both in networks and in the plane, is breaking symmetry.
To see why this is necessary, consider a highly symmetric network, such as an oriented ring (i.e., a ring in which ports at all nodes are labeled as follows:  
0 the clockwise port and 1 the counterclockwise port) or consider the infinite plane without obstacles.  Consider two identical agents starting at distinct nodes of the ring or at any two points in the plane, and running the same deterministic algorithm. It is easy to see that if they start simultaneously and move synchronously, they will never meet. In the ring, at all times they will use the port (at their respective current nodes) 
having the same label (as their history
is the same and the algorithm is deterministic), and hence the distance between them will be always the same. Likewise, in the plane, if the agents start simultaneously, have the same compass, the same measure of length and the same speed, they will traverse parallel trajectories and remain at the same distance at all times. 

In the deterministic scenario there are two ways of breaking symmetry. The first is by distinguishing the agents: each of them has a label and the labels are different.
Each agent knows its label, but we do not need to assume that it knows the label of the other agent. (If it does, then the solution is the well-known algorithm
Wait For Mommy: the agent with the smaller label stays idle, while the other one explores the graph or the plane in order to find it.) Both agents use the same {\em parametrized}
algorithm with the agent's label as the parameter. To see how this can help, consider two agents  that have to meet 
in an oriented ring of known size $n$. As mentioned above, if agents are anonymous (and marking nodes is disallowed), rendezvous is impossible. Now assume
that agents have distinct labels $L_1$ and $L_2$. A simple (although inefficient) rendezvous algorithm is: 
Make $L$ tours of the ring, where $L$ is your label, and stop.
Then the agent with larger label will make at least one full tour of the ring while the other one is already inert, thus guaranteeing rendezvous.

The second way of breaking symmetry, available even when agents are anonymous, is by exploiting either non-symmetries of the network itself, or the differences of the initial positions of the agents,
 even in a symmetric network. This method is impossible to use in the plane without obstacles, and in the network environment it is usable only for some classes
 of networks, as either the network must have distinguishable  nodes that play the role of ``focal points'' or the initial positions of agents have to be ``non-symmetric''  (the precise meaning of this condition will be defined later). 
 As a simple example of the application of this method, consider a $n$-node line with two identical agents. If $n$ is odd, then the line contains a central node that both agents
 can identify and meet at this node. If $n$ is even, (and even when the port labelings are symmetric with respect to the axis of symmetry of the line)  but the initial positions of the agents have different distances from their closest extremity, then the following algorithm works:
 Compute your distance $d$ from the closest extremity of the line, then traverse the line $d$ times and stop.
 For the same reasons as before, this algorithm guarantees rendezvous, whenever the initial positions of the agents are not symmetrically situated. On the other hand, if they are symmetric (and port labelings are symmetric as well), then it is easy to see that rendezvous is impossible if agents have to meet at a node and do not realize crossings on an edge. 
 
 We now discuss methodological implications of the distinction between the second pair of alternative assumptions, yielding the synchronous and asynchronous scenarios. The discussion is for agents with distinct labels. In the first case, the ability of the agents to exploit time, and more precisely, to vary between carefully measured periods of activity, when the agent explores parts of the network or of the plane, and of passivity, when it stays idle, is a powerful tool in the solution of the  rendezvous problem. Indeed, agents may exploit differences in their labels to schedule  these activity and passivity periods in such a way that at some point the active agent must 
 visit the position of the agent that is currently passive, and thus accomplish rendezvous. No such possibility is available in the asynchronous scenario. In this case, the main methodological tool is constructing trajectories of the agents, again exploiting the differences of their labels, in such a way that parts of these trajectories coincide, and  the agents are forced, regardless of their speed, to traverse a common segment of the trajectory at approximately the same time, implying rendezvous. It should be mentioned that, in the asynchronous scenario applied to networks,
  it may be impossible to meet at a node, and thus the requirement is relaxed to that of meeting at a node or inside an edge.
  
 In order to make the statement of a rendezvous problem precise, we have to point out what exactly is being sought, apart from meeting. The most general
question is that of feasibility: for what classes of networks and what initial positions is rendezvous possible under a particular scenario, and when is it possible in the plane? Here a complete solution
would be to prove that for some classes of networks and some initial configurations of agents rendezvous is impossible, and to provide a rendezvous algorithm for all other situations.
In the case of the plane, the question is whether approach starting from arbitrary unknown positions is always possible under a given scenario. 
More specific questions concern the amount of resources needed for rendezvous. These are usually of two types. The first is rendezvous {\em cost}:
the maximum number of steps made by an agent until rendezvous, or the maximum time used by the agents to meet. Algorithms minimizing the cost (or its order of magnitude) are sought in this context.
The other important resource is {\em memory}: what is the minimum memory with which agents have to be equipped in order to solve the rendezvous problem in a given class of networks. 
When only cost optimization is sought, memory of the agents is often assumed to be unbounded and they
are modeled as Turing machines. In problems seeking memory minimization (or tradeoffs between memory and time), the model of input/output automata 
(finite state machines) is usually used.
As usual in optimization tasks, a complete solution calls for an algorithm with given cost or memory and for an accompanying lower bound showing that this cost or amount of memory is optimal.

As mentioned before, in Section 3 we present results concerning rendezvous in networks, while in Section 4
we study the problem of approach in the plane.
The study is further subdivided by considering the
synchronous and asynchronous scenarios. Further assumptions are added when presenting a particular model.
In each case, we first give a precise description of the model, state the problem to be solved, then present the results  and often give a high-level description of methods and algorithms used to obtain them.

\section{Rendezvous in networks}

In this section we survey results on deterministic rendezvous in networks, dividing our considerations into two major scenarios : synchronous and asynchronous.

\subsection{Synchronous rendezvous}

Agents move in synchronous rounds. In every round, an agent may either remain at the same node or move to an 
adjacent node. Rendezvous means that all agents are at the same node in the same round.
Agents that cross each other when moving along the same edge, do not notice this fact.
Two subscenarios are considered: {\em simultaneous startup}, when both agents
start executing the algorithm in the same round, and {\em arbitrary delay}, when starting rounds are
arbitrarily decided by an adversary. In the former case, agents know that starting rounds are the same, while 
in the latter case, they are not aware of the difference between starting rounds,
and each of them starts executing the rendezvous algorithm and counting steps in the round of its own startup.

We will discuss separately the sub scenario of labeled agents, where agents have distinct integer labels that they can use as a parameter in the common deterministic algorithm, and that of anonymous agents, where agents do not have any labels and thus are identical.

\subsubsection{Labeled agents}

In \cite{DFKP} (whose journal version was published in 2006, but which is based on two earlier conference papers published in 2003 and 2004),   rendezvous of two agents is considered and it is indicated that all results can be generalized to an arbitrary number of agents.
It is assumed that  agents have different positive integer labels, and each agent knows its own label (which is a parameter
of  the common deterministic algorithm that they use), but is unaware of the label of the other agent. 
In general, agents do not know the topology of the graph in which they have to meet.  It is assumed that the agents have unlimited memory
(they are modeled as Turing machines)
and the authors aim at optimizing the cost of rendezvous.
This cost is defined as the worst-case number of rounds since the startup of the later 
agent until rendezvous is achieved, where the worst case is taken over all graphs in the considered class, all initial positions of the agents
and all possible startup times (decided by an adversary), in the case of the arbitrary delay scenario.

The following notation is used. The labels of the agents are $L_1$ and $L_2$.
The smaller of the two labels is denoted by $l$. The delay (the difference between startup times of the
agents) is denoted by $\tau$,  $n$ denotes the number of nodes in the graph,
and $D$ --  the distance between initial positions of agents.

The authors  introduce the problem in the relatively simple case of  trees. They show that
rendezvous can be completed
at cost $O(n + \log l)$ on any $n$-node tree, 
even with arbitrary delay. 
They also show that
for some trees this complexity cannot be improved, even with simultaneous startup. 
Rendezvous in trees is relatively easy for two reasons. First, a tree can be explored with termination and a map of it can be constructed by a single agent,
using the {\em basic walk} which consists in leaving every node by the next port with respect to the entry port at this node (modulo the degree). The second reason is the
existence of the central node or the central edge in any tree. Once each agent locates this object independently, the central node plays the role of the ``focal point''
and rendezvous can be accomplished at linear cost. If there exists a central edge, rendezvous is slightly more complicated and, after identifying this edge,
it reduces to rendezvous in the two-node graph. It is this case that is responsible for the  $O( \log l)$ additive term in the cost complexity.

As soon as the graph contains cycles, another technique has to be applied. The authors
continue the study by concentrating on the simplest class of such graphs, i.e., rings.
They prove that, with simultaneous startup,
the optimal cost of rendezvous on any ring is $\Theta(D \log l)$. 
They construct an algorithm achieving rendezvous with this complexity and show that, for any distance $D$, it 
cannot be improved. 

The lower bound $\Omega(D \log l)$ relies on the following idea. The (oriented) ring is partitioned into pieces of equal size $\Theta(D)$ and time is partitioned into segments
of the same length. It is observed that at the end of a segment the agent can be either in the same piece as in the beginning of it, or in one of the neighboring pieces.
This permits to code the behavior of an agent by a ternary sequence corresponding to its position at the end of each time segment. It is argued that if two agents have the same behavior code, then they cannot meet. Moreover, if the time of rendezvous were   $o(D \log l)$, then behavior codes would have to be short, and thus for some two different labels
of agents they would have to be the same, as the behavior of an agent before the meeting depends only on its label.
Assigning these labels to the agents would preclude rendezvous.

With an arbitrary delay, $\Omega(n+ D \log l)$ is a lower bound on the  
cost required for rendezvous in a $n$-node ring.
Under this scenario,  two rendezvous algorithms for
the ring are presented in \cite{DFKP}: an algorithm with cost $O(n \log l)$, for known $n$, and an algorithm with cost
$O(l \tau + ln^2)$,  if $n$ is unknown. In the latter case, the cost was later improved to $O(n \log l)$ in \cite{KM}.
In view of the above lower bound, this  cannot be improved in general.

For arbitrary connected graphs, the main contribution of  \cite{DFKP} is the first deterministic rendezvous algorithm with cost
polynomial in $n$, $\tau$ and $\log l$. More precisely, the authors present an algorithm that
solves the rendezvous problem for any $n$-node graph $G$, for
any labels $L_1>L_2=l$ of agents and for any delay $\tau$ between startup times,
in cost $\cO(n^5\sqrt{\tau\log l}\log n + n^{10}\log^2 n \log l)$.
The algorithm contains a non-constructive
ingredient: agents use combinatorial objects whose existence is proved by the
probabilistic method.
Nevertheless the algorithm is indeed deterministic. Both agents can find
separately
the same combinatorial object with the desired properties (which is then used
in the rendezvous algorithm). This can be done
using brute force exhaustive search which may be quite complex but in the adopted model
only moves of the agents are counted and local
computation time of the agents does not contribute to cost.
Finally, the authors prove a lower bound $\Omega (n^2)$ on the cost of rendezvous in some  
family of graphs.   
%


The paper is concluded by an open problem concerning
the dependence of rendezvous cost on the delay
$\tau$. The dependence on the
other parameters follows from the results cited above. Indeed,
 a lower bound $\Omega (n^2)$ on rendezvous cost has been shown in
some graphs. The authors also showed that cost $\Omega (\log l)$ is required
even for the two-node graph.
On the other hand, for agents starting at distance
$\Omega(n)$ in a ring, cost $\Omega(n\log l)$ is required,
even for $\tau =0$.
However, since the complexity of their algorithm contains a factor $\sqrt{\tau}$, the authors state the following problem:

Does there exist a deterministic rendezvous algorithm for arbitrary connected graphs with cost polynomial
in $n$ and $l$ (or even in $n$ and $\log l$) but independent of $\tau$? 

A positive answer to this problem was  given in \cite{KM} (whose conference version was published in 2006).
The authors present a rendezvous algorithm for two agents, working in arbitrary connected graphs for an arbitrary delay $\tau$, 
whose complexity is $O(\log^3 l +n^{15}\log ^{12} n)$, i.e., is independent 
of  $\tau$ and polynomial in  $n$ and $\log l$. As before, the algorithm contains a non-constructive ingredient, but is deterministic.

The rendezvous algorithms from \cite{DFKP,KM}, working for arbitrary connected graphs, yield an intriguing question, stated in  \cite{DFKP}. 
While both of them have polynomial cost
(the one from \cite{DFKP} depending on $\tau$, and the one from \cite{KM} independent of $\tau$), they both use a non-constructive ingredient, i.e, 
a combinatorial object whose existence is proved using the probabilistic method. As mentioned above, each of the agents can deterministically find such an object
by exhaustive search, and then use it in the execution of its algorithm, which keeps the algorithm deterministic, but may significantly increase the time of local computations. In the described model the time of these local computations does not contribute to cost which is measured by the number of steps, regardless of the time taken to compute each step. Nevertheless, it is interesting if there exists a rendezvous algorithm for which both the cost and the time of local computations are
polynomial in $n$ and $\log l$. Such an algorithm would have to eliminate any non-constructive ingredients.

This problem was solved in \cite{TZ} (whose conference version appeared in 2007), using the important notion of a
 {\em Universal Exploration Sequence} (UXS)  \cite{koucky01b}.
Let $(a_1,a_2,\dots, a_k)$ be a sequence of  integers. An {\em application} of this sequence to a graph $G$ at node $u$
is the sequence of nodes $(u_0,\dots , u_{k+1})$ obtained as follows: $u_0=u$, $u_1$ is the node joined to $u$ by the edge corresponding
to port 0 at $u$; for any  $1 \leq i \leq k$, $u_{i+1}$ is the node joined to $u_i$ by the edge corresponding (at $u_i$) to port 
$(p+a_i) \mod d(u_i)$, where $p$ is the port number
 at $u_i$ corresponding to the edge $\{u_i,u_{i-1}\}$ and $d(u_i)$ denotes the degree of node $u_i$. 
(Informally, an application of  $(a_1,a_2,\dots, a_k)$ corresponds to a walk in the graph in which the current exit port is computed by adding $a_i$ to the 
current entry port.)
 A sequence $(a_1,a_2,\dots, a_k)$ whose application
 to a graph $G$ at any node $u$ contains all nodes of this graph is called a UXS for this graph.
A UXS for a class $\cG$ of
graphs is a UXS for all graphs in this class. The solution from \cite{TZ} uses a result following from \cite{R08} that a UXS for the class of all connected graphs
with at most $n$ nodes can be computed in time polynomial in $n$. Moreover, the authors propose a rendezvous algorithm working in $O(n^5 \log l)$ rounds 
(up to factors logarithmic in $n$ and $\log l$). This complexity beats those from \cite{DFKP,KM} and makes the algorithm from \cite{TZ} the currently most efficient rendezvous algorithm working in arbitrary connected graphs.

The paper \cite{DPP} started the investigation of gathering of a team of labeled agents, some of which can be Byzantine.
The size of the team is unknown to the agents. Agents can exchange all currently held information when they meet at a node of the graph.
Up to $f$ of the agents are Byzantine. The authors define two levels of Byzantine behavior.
A strongly Byzantine agent can choose an arbitrary port when it moves and it can transmit arbitrary
information to other agents, while a weakly Byzantine agent can do the same, except changing its label. The main problem investigated in the paper is what is the minimum number of good agents that
guarantees deterministic gathering of all of them, with termination. (Of course, Byzantine agents cannot be forced to gather.)  The authors solve exactly this Byzantine gathering problem in arbitrary
networks for weakly Byzantine agents, and give approximate solutions for strongly Byzantine agents, both when the size of the network is known
and when it is unknown. They show that both the strength versus weakness of Byzantine behavior and the knowledge of network size have an important influence on the results.

For weakly Byzantine agents it is shown that any number of  good agents permit to solve the problem for networks of known size.
If the size is unknown, then this minimum number is $f+2$. More precisely, the authors design a deterministic polynomial algorithm 
that gathers all good agents in an arbitrary network, provided that there are at least $f+2$ of them. They also prove a matching lower bound showing that
 if the number of good agents is at most $f+1$, then they are not able to gather deterministically with termination in some networks.

For strongly Byzantine agents the authors give a lower bound of $f+1$, even when the graph is known: they show that  $f$ good agents cannot gather deterministically
in the presence of $f$ Byzantine agents even in a ring of known size. In order to establish upper bounds, they propose deterministic gathering algorithms for at least $2f+1$ good agents when the size of the network is known,
and for  at least $4f+2$ good agents when it is unknown. These upper bounds were subsequently improved in \cite{BDD} to $f+1$ when the size of the network is known and to $f+2$ when it is unknown. Together with the lower bounds from \cite{DPP}, both these upper bounds are tight. 

As the authors of \cite{BDD} point out, the above results
show an interesting difference between the scenarios of known vs. unknown size of the network.
While for known size, the gap between the number of good agents permitting gathering with weakly and with strongly Byzantine agents
is very significant (1 vs. $f+1$) this gap completely disappears for the scenario of unknown size: the minimum number of good agents is then $f+2$, regardless of whether the bad agents
are weakly or strongly Byzantine.

Gathering of agents in the presence of a  more benign type of faults is considered in \cite{Pe2}. In this paper it is assumed that some agents are subject to
 {\em crash faults} which can occur at any time.
Two fault scenarios are considered. A {\em motion fault} immobilizes the agent at a node or inside an edge but leaves intact its memory at the time
when the fault occurred. A more severe {\em total fault} immobilizes the agent as well, but also erases its entire memory. As before, we cannot require faulty agents to gather.
Thus the gathering problem for fault prone agents calls for all fault-free agents to gather at a single node, and terminate.  

It is observed that when agents move completely asynchronously, gathering with crash faults of any type is impossible. Hence the author considers a restricted version of asynchrony, where each agent is assigned by the adversary a fixed speed, possibly different for each agent. Agents have clocks ticking at the same rate. Each agent can wait for a time
of its choice at any node, or decide to traverse an edge but then it moves at constant speed assigned to it. It is appropriate to discuss this model in the section devoted to synchronous algorithms, as methodologically the two scenarios are similar: it is still possible to wait at a node for a prescribed amount of time, and this capability significantly influences algorithm design. When two or more agents are at the same node or in the same point of an edge in the same time, they can see the memory content of other agents at this node or at this point of an edge, except for memory of faulty agents in the case of  total faults.

The main results of the paper are a gathering algorithm working for any team of at least two agents in the scenario of motion faults, and a gathering algorithm working in the presence of total faults, provided that at least two agents are fault free all the time. If only one agent is fault free, the task of gathering with total faults is sometimes impossible. This shows that in the case of crash faults more faulty agents can be tolerated for gathering than when faults are Byzantine. Both algorithms from \cite{Pe2} work in time
polynomial in the size of the graph, in the logarithm of the largest label, in the inverse of the smallest speed, and in the ratio between the largest and the smallest speed. 

Rendezvous of two agents subject to transient faults is considered in \cite{CDLP}.
Agents do not know the topology of the network or any bound on its size. 
In each round an agent decides if it remains
idle or if it wants to move to one of the adjacent nodes. Agents are subject to {\em delay faults}: if an agent incurs a fault in a given round,
it remains in the current node, regardless of its decision. If it planned to move and the fault happened, the agent is aware of it. The authors consider three
scenarios of fault distribution: random (delay faults occur independently in each round and for each agent with constant probability $0<p<1$), unbounded 
adversarial (the adversary can delay an agent for an arbitrary finite number of consecutive rounds) and bounded adversarial
(the adversary can delay an agent for at most $c$ consecutive rounds, where $c$ is unknown to the agents). The quality measure of a rendezvous algorithm is its cost, which is the total number of edge traversals.

For random faults, the authors design an algorithm with cost polynomial in the size $n$ of the network and polylogarithmic in the larger label $L$, which achieves rendezvous with probability at least $1-1/n$ in arbitrary networks.
By contrast, for unbounded adversarial faults they prove that rendezvous is not feasible, even in the class of rings.
Under this scenario, the authors design a rendezvous algorithm with cost $O(n\ell)$, where $\ell$ is the smaller label,
working in arbitrary trees, and they show that $\Omega(\ell)$ is the lower bound on rendezvous cost, even for the two-node tree.
For bounded adversarial faults, the authors construct a rendezvous algorithm working  for arbitrary networks, with cost polynomial in $n$,
and logarithmic in the bound $c$ and in the larger
label $L$. 

In \cite{ODM}, the authors use a different approach to counter transient faults. They propose a {\em self-stabilizing} rendezvous algorithm for two agents navigating in an arbitrary network. A self-stabilizing algorithm has the property that if the agents start from any two nodes with arbitrary memory states, then eventually they will get to the same node in the same round. Thus, even if the agents incur a transient fault of any kind that corrupts their memory, they can meet in finite time after the fault disappears.
The authors design a self-stabilizing rendezvous algorithm for arbitrary graphs, without any time guarantees, and construct polynomial-time self-stabilizing rendezvous algorithms for trees and rings. More precisely, the algorithm for trees is polynomial in the size of the tree and in the logarithm of the smaller label, and the algorithm for rings is polynomial in the size of the ring and in both labels.

In \cite{MP}, the authors consider 
two main efficiency measures of  rendezvous:  its  time, i.e., the number of rounds until the meeting,
and its cost, i.e., the total number of edge traversals. They investigate tradeoffs between these two measures.
A natural benchmark for both time and cost of rendezvous in a network is the number of edge traversals needed for visiting all nodes of the network,
called the exploration time. The authors express the time and cost 
of rendezvous as functions of an upper bound $E$ on the time of exploration (where $E$ and a corresponding exploration procedure are known to both agents) and of the size $L$ of the label space.
They design two rendezvous algorithms:  algorithm {\tt Cheap} has cost $O(E)$ and time $O(EL)$, and  algorithm {\tt Fast} has both time and cost $O(E\log L)$. The main contribution of the paper are lower bounds showing that
these two algorithms capture the tradeoffs between time and cost of rendezvous almost tightly. They show that
any deterministic rendezvous algorithm of cost asymptotically $E$ (i.e., of cost $E+o(E)$) must have time $\Omega(EL)$. On the other hand,
they prove that any deterministic rendezvous algorithm with time complexity $O(E\log L)$ must have cost $\Omega (E\log L)$.

In \cite{MP2}, the problem of rendezvous is studied in the framework of {\em advice}, which is a popular paradigm permitting to measure the amount of information that agents need in order to perform some task in networks.
If $D$ is the distance between the initial positions of the agents, then $\Omega(D)$ is a lower bound on the time of rendezvous. However, in the absence of any knowledge about the network, agents usually cannot meet in time 
$O(D)$. Thus the authors study the minimum amount of information that has to be available  {\em a priori} to the agents in order to achieve rendezvous in optimal time $\Theta(D)$.
Following the advice paradigm, this information is provided to the agents at the start by an oracle knowing the entire instance of the problem, i.e., the network, the starting  positions of the agents, their wake-up rounds,  and both of their labels. The oracle provides
the agents with the same binary string called {\em advice}, which can be used by the agents during their rendezvous algorithm. The length of this
string is called the {\em size of advice}.  The goal of the paper is to find the smallest size of advice which enables the agents to meet in time $\Theta(D)$. 
The authors solve this problem completely by showing that this optimal size of advice is $\Theta(D\log(n/D)+\log\log L)$,
where $n$ is the size of the graph, $D$ is the initial distance between the agents, and $L$ is the size of the label space. 
The upper bound is proved by constructing an advice string of this size, and providing a rendezvous algorithm using this advice that works in time $\Theta(D)$
for all networks.
The matching lower bound, which is the most difficult and interesting part of the paper, is proved by constructing classes of networks for which it is impossible to achieve
rendezvous in time $\Theta(D)$
with smaller advice. 

The authors of \cite{DDKU} investigate the rendezvous problem in graphs under the scenario where during navigation each agent gets some restricted feedback about the position of the other agent. More precisely, they consider distance-aware agents that, in every round, are informed of the distance between them. The authors show that such agents can meet in time $O(\Delta (D +\log l))$,
where $D$ is the initial distance between the two agents, $l$ is the smaller label and $\Delta$ is the maximum degree of the graph. Thus, even in a very large graph, distance-aware agents can meet in time polynomial in local parameters of the instance of the rendezvous problem. Moreover, the authors show an almost matching lower bound $\Omega(\Delta (D +\log l/\log \Delta))$ on the time of rendezvous in their scenario. 

In most formulations of the synchronous rendezvous problem,
meeting is accomplished when the agents get to the same node in the same round. In \cite{EP}, the authors consider a more demanding task, called {\em rendezvous with detection}: agents must become aware that the meeting is accomplished,
simultaneously declare this and stop. It is clear that in order to signal to the other agent the presence at a given node, agents must communicate, and the awareness of the meeting depends on ways of communication between the agents. The authors study two variations of a very weak model of communication, called the {\em beeping model}, introduced in \cite{CK}. In each round an agent can either listen or beep. In the {\em local beeping model}, an agent hears a beep in a round if it listens in this round and if the other agent is at the same node and beeps. In the {\em global beeping model},
an agent hears a {\em loud} beep in a round if it listens in this round and if the other agent is at the same node and beeps, and it hears a {\em soft}
beep in a round if it listens in this round and if the other agent is at some other node and beeps.

The authors propose a deterministic algorithm of rendezvous with detection working, even for the weaker local beeping model, in an arbitrary unknown network in time polynomial in the size of the network and in the logarithm of the smaller label. However, this algorithm is highly energy consuming: the number of moves that an agent must make, is proportional to the time of rendezvous. Hence the authors ask if {\em bounded-energy agents}, i.e., agents that can make at most $c$ moves, for some integer $c$, can always achieve rendezvous with detection as well.
They observe that this is impossible for some networks of unbounded size. Hence they rephrase the question as follows. Can bounded-energy agents always achieve rendezvous with detection in bounded-size networks? The authors prove that the answer to this question is positive, even in the local beeping model but  this ability comes at a steep price of time: the meeting time of bounded-energy agents is exponentially larger
than that of unrestricted agents. By contrast, the authors propose an algorithm for rendezvous with detection in the global beeping model that works for bounded-energy agents
(in bounded-size networks) as fast as for unrestricted agents.

We conclude this section by discussing rendezvous of agents that have fixed but possibly different speeds. The meeting must be at a node, which precludes the fully asynchronous scenario.
Hence, the authors of \cite{BDPP} consider a scenario of
agents with restricted asynchrony: agents have the same measure of time but the adversary can arbitrarily impose the speed of traversing each edge by each of the agents. They construct a rendezvous algorithm for such agents, working in time polynomial in the size of the graph, in the length of the smaller label, and in the largest edge traversal time.

\subsubsection{Anonymous agents}

One of the first papers on synchronous rendezvous of two anonymous agents was \cite{EP3} where the authors compare the time of deterministic and of randomized rendezvous in trees.
For deterministic rendezvous they propose an algorithm working in time linear in the size of the tree, for every initial configuration for which rendezvous is possible, and they show that this time cannot be improved in general, even when agents start at distance 1 in bounded degree trees.

In the case of anonymous agents, rendezvous may be impossible for some initial configurations, in some networks, as witnessed by the example of two agents in the oriented ring, mentioned before. This yields an important feasibility problem, which is to characterize those initial configurations of an arbitrary number of anonymous agents for which rendezvous (gathering) is feasible, and to provide a gathering algorithm working for all such configurations. This problem was attacked and completely solved in 
\cite{DP}.

At least two agents start from different nodes of the graph.
The adversary wakes up some of the agents at possibly different times. A dormant agent, not woken up by the adversary,  is woken up by the first agent that visits
its starting node, if such an agent exists. 
Agents do not  know the topology of the graph or the size of the team. The authors considered two scenarios: one when agents know
a polynomial upper bound on the size of the graph and another when no bound is known. 
When several agents are at the same node in the same round,
they  can exchange all information they currently have. 
The authors  assume that the memory of the agents is unlimited.

An initial configuration of agents, i.e., their placement at some distinct nodes of the graph, is called {\em gatherable} if there exists a deterministic algorithm 
(even only dedicated to this particular configuration) that achieves meeting of all
agents in one node, regardless of the times at which some of the agents are woken up by the adversary. The authors study the problem of
which initial configurations are gatherable and how to gather all of them deterministically by the same algorithm.
The problem calls for deciding which initial configurations are possible to gather, even by an algorithm specifically designed for this particular
configuration, and for finding a {\em universal} gathering algorithm that gathers all such configurations. The authors restrict attention only to {\em terminating}
algorithms, in which every agent eventually stops forever.

The main result of the paper is a complete solution of the above gathering problem in arbitrary graphs. The authors characterize all gatherable configurations and give two {\em universal} deterministic 
gathering algorithms, i.e., algorithms that gather all gatherable configurations. The first algorithm works under the assumption that an upper bound $n$
on the size of the network is known. In this case their algorithm guarantees {\em gathering with detection}, i.e., the existence of a round in which,
for any gatherable configuration,  all agents are at the
same node and all declare that gathering is accomplished. If no upper bound on the size of the network is known, the authors show that a universal algorithm for gathering
with detection does not exist. Hence, for this harder scenario, they construct a second universal gathering algorithm, which guarantees that, for any gatherable
configuration, all agents eventually get to one node and stop, although they cannot tell if gathering is over. The time of the first algorithm is polynomial in the
upper bound $n$ on the size of the network, and the time of the second algorithm is polynomial in the (unknown) size of the network itself.

As pointed out by the authors of \cite{DP}, the problem of gathering an unknown team of anonymous agents in an arbitrary network  
presents the following major difficulty. The asymmetry of the initial configuration
because of which gathering is feasible, may be caused not only by non-similar locations
of the agents with respect to the structure of the graph, but by their different situation with respect to other agents.
Hence the authors had to come up with a new algorithmic idea: in order to gather, agents
that were initially identical, must make decisions based on the memories
of other agents met to date, in order to make their future behavior different and break symmetry in this way.
In the beginning the memory of each agent is empty and
in the execution of the algorithm it records what the agent has seen in 
previous steps of the navigation and what it heard from other agents that it met.
Even small differences occurring  in a remote part of the graph
must eventually influence the behavior of initially distant agents.
Agents in different initial situations may be unaware of this difference in early meetings, as the difference
may depend on their location with respect to remote agents and thus be revealed only later on. Hence an agent may mistakenly consider that two different agents
that it met in different stages of the algorithm execution, are the same agent. 
Confusions due to this possibility are a significant
challenge in the algorithm design,  that occurs neither in gathering two (even anonymous) agents nor in gathering many labeled agents.

Rendezvous of two anonymous agents was considered in \cite{CKP,FP2}. 
As mentioned before,  in this case rendezvous is not possible for arbitrary networks and arbitrary initial positions of the agents.
In order to describe initial positions of the agents for which rendezvous is possible, we need the notion of a {\em view} from a node of a graph, introduced in  \cite{YK3}. Let $G$ be a graph and $v$ a node of $G$. The {\em view} from $v$ is the infinite tree $\cV(v)$ rooted at $v$ with labeled
ports, whose branches are infinite paths in $G$ starting at $v$, coded as sequences of ports.
A pair $(u,v)$ of distinct nodes is called {\em symmetric}, if $\cV(u)=\cV(v)$. Initial positions forming a symmetric pair
of nodes are crucial when considering the feasibility of rendezvous in arbitrary graphs. Indeed, it is proved in \cite{CKP} that rendezvous is feasible, if and only if
the initial positions of the agents are not a symmetric pair. For the particular case of the class of trees, this is equivalent to the non-existence of a port-preserving
automorphism of the tree that carries one initial position to the other.

The aim of \cite{CKP,FP2} was optimizing the memory size of the agents that seek rendezvous.
In order to model agents with bounded memory, the formalism of input/output automata is used.
An agent is an abstract state machine
$\cA=(S,\pi,\lambda,s_0)$, where $S$ is a set of states among which there is
a specified state $s_0$ called the {\em initial} state, $\pi:S\times
\mathbb{Z}^2 \to S$, and $\lambda:S\to \mathbb{Z}$. Initially the agent is at some node
$u_0$, called its {\em initial position}, in the initial state $s_0\in S$. The agent performs an action
in each step. Each action can be either a move to an adjacent
node or a null move resulting in remaining in the currently occupied node.
State $s_0$ determines an integer
number $\lambda(s_0)$. If $\lambda(s_0)=-1$, then the agent makes a null move (i.e., remains at $u_0$).
If $\lambda(s_0)\geq 0$ then the agent leaves $u_0$ by port $\lambda(s_0)$ mod $d(u_0)$, where $d(u_0)$ is the degree of $u_0$ . When
incoming to a node $v$ in state $s\in S$, the behavior of the agent is as follows.
It reads the number $i$ of the port through which it entered $v$ and
the degree $d$ of $v$. The pair $(i,d)\in \mathbb{Z}^2$ is an input symbol
that causes the transition from state $s$ to state
$s'=\pi(s,(i,d))$. If the previous move of the agent was null,
(i.e., the agent stayed at node $v$ in state $s$), then the pair $(-1,d)\in \mathbb{Z}^2$ is the input symbol
read by the agent, that causes the transition from state $s$ to state
$s'=\pi(s,(-1,d))$.
In both cases $s'$ determines an integer
$\lambda(s')$, which is either $-1$, in which case the agent makes a null move, or
a non negative integer indicating the port number
by which the agent leaves $v$. This port number is $\lambda (s')$ mod $d(v)$, where $d(v)$ is the degree of $v$. The agent continues
moving in this way, possibly infinitely.

In order to grasp the assumption that agents are identical, it is assumed that agents are
copies $A$ and $A'$ of the same abstract state machine $\cA$, starting at two distinct nodes $v_A$ and $v_{A'}$.
We will refer to such identical machines as a {\em pair of agents}. 
A pair of agents is said to solve the rendezvous problem {\em  with delay} $\tau$
in a class $\cC$  of graphs, if, for any graph in the class $\cC$
and for any initial positions that are not symmetric, both agents are eventually in the same node of the graph in the same round,
provided that they start with delay $\tau$. The memory of an agent is measured by the number of states of the corresponding automaton, or equivalently by the number of bits on which these states are encoded. An automaton with $K$ states requires $\Theta (\log K)$ bits of memory.

In \cite{FP2} (based on conference papers published by the same authors in DISC 2008 and SPAA 2010) the authors focus attention on optimizing memory size of identical agents that permits them meeting in trees. 
They assume that the port labeling is decided by an adversary aiming at preventing two agents from meeting, or at allowing the agents to meet only after having consumed a lot of resources, e.g., memory space. This yields the following definition.
A pair of agents initially placed at  nodes $u$ and $v$ of 
a tree $T$ solves the rendezvous problem if, for any port labeling of 
$T$, both agents are eventually in the same node of the tree in the 
same round. 

Nodes $u$ and $v$ of a tree $T=(V,E)$ are {\em perfectly symmetrizable} if there exists a port labeling $\mu$ of $T$
and an automorphism of the tree preserving $\mu$
that carries one node on the other. 
According to the above definition, the condition on  feasibility of rendezvous can be reformulated as follows: 
a pair of agents can solve the rendezvous problem in a tree, if and only if their initial positions are not perfectly symmetrizable. 
Consequently, throughout \cite{FP2}, the authors consider only non perfectly symmetrizable initial positions of the agents.

It is first shown that the minimum size of memory of the agents that can solve the rendezvous problem in the class of trees with at most $n$ nodes is $\Theta (\log n)$. A rendezvous algorithm for arbitrary delay $\tau$,
that uses only a logarithmic number of memory bits follows, e.g., from  \cite{CKP}.  It is observed in \cite{FP2} that $\Omega (\log n)$ is also a lower bound on the number of bits of memory
that permit the agents to meet in all trees of size linear in $n$. 
Due to this lower bound, a {\em universal} pair of finite agents achieving rendezvous in the class of all trees cannot exist.
However, the lower bound uses a counterexample of a tree with maximum degree linear in the size of the tree. 
Hence, it is natural to ask if there exists a pair of finite agents solving the rendezvous problem in all trees of {\em bounded} degree.
The authors give a negative answer to this question. In fact they show that, for any pair of identical finite agents, there is a line for which these agents cannot solve the 
rendezvous problem, even with simultaneous start. As a function of the size of the trees, this impossibility result indicates a lower bound $\Omega(\log\log n)$ bits on the memory size for rendezvous in bounded degree trees of at most $n$ nodes. 

The main topic of  \cite{FP2} is the impact of the delay between startup times of agents on the minimum size of memory permitting rendezvous.
The authors show that if this delay
is arbitrary, then the lower bound on memory required for rendezvous is $\Omega (\log n)$ bits,
even for the line of length $n$. This lower bound matches the upper bound from \cite{CKP}, which shows that 
the minimum size of memory of the agents that can solve the rendezvous problem in the class of 
{\em bounded degree} trees with at most $n$ nodes is $\Theta (\log n)$.
By contrast, for simultaneous start,  they show that
the amount of memory needed for rendezvous depends on two parameters of the tree: the number
$n$ of nodes and the number $\ell$ of leaves. Indeed, they construct two identical agents with $O(\log \ell + \log\log n)$
bits of memory that solve the rendezvous problem in all trees with $n$ nodes and $\ell$ leaves.
For the class of trees with $O(\log n)$ leaves, this proves an exponential gap in minimum
memory size of the agents permitting them to meet, between the scenario
with arbitrary delay and with delay zero.

Moreover, it is shown in \cite{FP2} that the size $\Theta(\log \ell + \log\log n)$ of memory used to solve the rendezvous problem
with simultaneous start
in trees with at most $n$ nodes and at most $\ell$ leaves is optimal, even in the class of trees with degrees bounded
by 3. More precisely,  for infinitely many integers $\ell$,  the authors show a class of arbitrarily large trees with maximum degree 3
and with $\ell$ leaves, for which rendezvous with simultaneous start requires $\Omega (\log \ell)$ bits of memory. This lower bound, together
with the previously mentioned lower bound $\Omega(\log \log n )$ on the number of bits of memory required to meet with simultaneous start in the line of length $n$,  
implies that the upper bound $O(\log \ell + \log\log n)$ cannot be improved even for trees with maximum degree at most 3.

Trade-offs between the size of memory and the time of rendezvous in trees by identical agents are investigated in \cite{CKP2}.
The authors consider trees with a given port labeling and assume that there is no port-preserving automorphism of the tree that carries the initial
position of one agent to that of the other (otherwise rendezvous with simultaneous start is impossible).
The main result of the paper
is a tight trade-off between optimal time of rendezvous and the size of memory of the agents.
For agents with $k$ memory bits, it is shown that optimal rendezvous time is
$\Theta(n+n^2/k)$ in $n$-node trees. More precisely, if  $k \geq c\log n$, for some constant $c$,
the authors construct agents accomplishing rendezvous in arbitrary trees of unknown size $n$ in time $O(n+n^2/k)$, starting with arbitrary delay.
They also show that no pair of agents can accomplish rendezvous in time $o(n+n^2/k)$, even in the class of lines and even with
simultaneous start.

Trade-offs between the size of memory of the agents and the time of  rendezvous in trees are investigated in \cite{BIOKM} in a slightly different scenario:
the difference is in the definition of rendezvous.
The authors consider the rendezvous problem of any number of anonymous agents. To handle the case of symmetric trees they weaken the rendezvous requirements:
agents have to meet at one node if the tree is not symmetric, and at two neighboring nodes otherwise. In this latter case, some of the agents may finish the algorithm in one of the two nodes and other agents in the other node.  The authors observe that $\Omega(n)$ is a lower bound
on the time of rendezvous in the class of $n$-node trees and show that any algorithm achieving rendezvous in optimal (i.e., $O(n)$) time must use
$\Omega(n)$ bits of memory for each agent. Then they show a rendezvous algorithm that uses $O(n)$ time and $O(n)$ bits of memory per agent.
Finally they design a polynomial time algorithm using $O(\log n)$ bits of memory per agent. An additional feature of the algorithms from  \cite{BIOKM} is that
they can also work in an asynchronous setting: each agent independently identifies the node or one of the two nodes where meeting should occur, it reaches this node and stops.

While  \cite{FP2} solves the problem of minimum memory size needed for rendezvous in trees, the same problem for the class of arbitrary 
 graphs is solved in 
\cite{CKP}.  The authors establish the minimum size of the memory of agents that guarantees deterministic rendezvous when it is feasible.
They show that this minimum size is $\Theta(\log n)$, where $n$ is the size of the graph, regardless of the delay  between
the startup rounds of the agents. More precisely, the authors construct identical agents equipped with $\Theta(\log n)$ memory bits that solve the rendezvous
problem in all graphs with at most  $n$ nodes, when starting with any delay, 
and they prove a matching lower bound $\Omega(\log n)$ on the number of memory bits
needed to achieve rendezvous, even for simultaneous start. In fact, this lower bound is valid already on the class of rings. 

The positive result from  \cite{CKP}  is based on a result from \cite{R08} which implies that a (usually non-simple) path traversing all nodes
can be computed (node by node) in memory $O(\log n)$, for any graph with at most $n$ nodes.
Moreover,  logarithmic memory suffices to walk back and forth on this path. More precisely the result from  \cite{R08} states that,
for any positive integer $n$, there exists a UXS $Y(n)=(a_1,a_2,\dots, a_M)$ for the class $\cG_n$ of all graphs
with at most $n$ nodes, such that
$M$ is polynomial in $n$,
and for any $i \leq M$, the integer $a_i$ can be constructed using $O(\log n)$ bits of memory.

At a high level, the idea of the algorithm from  \cite{CKP} is the following. The authors introduce the notion of the {\em signature} $S(u)$ of node $u$ corresponding
to a given UXS. This is the sequence of entry and exit ports which are traversed by an application of the UXS at $u$. They show that if $Y$ is
a UXS  for the class $\cG_{n^2+n}$ of all graphs with at most $n^2+n$ nodes, and $S(u)$ denotes the signature of $u$ in a $n$-node graph $G$, corresponding
to the UXS $Y$, then for any nodes $v$, $w$ of $G$,  $\cV(v)\neq \cV(w)$ is equivalent to  $S(v)\neq S(w)$. Thus, the sequence $S(u)$ can be treated as a compact representation of the view $\cV(u)$. Then the authors show that, using logarithmic memory, it is possible to further compress $S(u)$ to a positive integer value of at most $n$ in such a way that different signatures correspond to different values.  This numerical value is then used as the label of the agent. Hence  
agents starting from non-symmetric initial positions get different labels. Once the agents' anonymity is broken, the rest of the meeting procedure is performed in the usual way,
by dividing time into segments corresponding to activity/passivity phases. Time segments are long enough to perform 
a complete exploration of $G$ (using a UXS that can be  computed in logarithmic memory). An agent explores $G$ in a single phase allowed for its label and waits in the remaining phases. Hence an agent performing its exploration phase must meet any agent of different label which is inert during this phase.

\subsection{Asynchronous rendezvous}

In the asynchronous scenario agents no longer perform their moves in synchronized steps. While the agent chooses the adjacent node to which it wants to go,
the time at which this move is executed and the possibly varying speed are chosen by an adversary, which considerably complicates rendezvous.  It is easy to see that,
even in the two-node graph, meeting at a node cannot be guaranteed under this scenario, hence the rendezvous requirement is relaxed by demanding only that
meeting occur either at a node or inside an edge.
Since  meetings
inside an edge are allowed,  
unwanted crossings of edges have to be avoided. Thus, for the asynchronous scenario, an embedding of the
underlying graph in the
three-dimensional Euclidean space is considered, with nodes of the graph being points of the
space and edges being
pairwise disjoint line segments joining them. For any graph such an embedding
exists. Agents are modeled
as points moving inside this embedding.

At any currently visited node, an agent executing a rendezvous algorithm chooses a port number at this node, corresponding to the edge that the agent wants to traverse.
However, the way of traversing this edge is decided by the adversary,
capturing the asynchronous characteristics of the rendezvous process.
When the agent, situated at a node $v$ at time $t_0$ has to traverse an edge
modeled as a segment $[v,w]$,
the adversary performs the following choice. It selects a time point $t_1>t_0$
and any continuous function
$f:[t_0,t_1] \longrightarrow [v,w]$, with $f(t_0)=v$ and $f(t_1)=w$. This
function models the actual movement of
the agent inside the
line segment $[v,w]$ in the time period $[t_0,t_1]$. Hence this movement can be
at arbitrary speed, the agent
may be even forced by the adversary to go back and forth, as long as it does not leave the segment and the movement
is continuous. 
We say that at time $t \in [t_0,t_1]$ the agent is in point $f(t) \in [v,w]$.
Moreover, the adversary 
chooses the starting time of the agent. Hence an agent's trajectory is
represented by the 
concatenation of the functions chosen by the adversary for consecutive edges
that the agent traverses.
Recall that the choice of the edge incident to a current node is determined by the choice of the port number, belonging to the agent.

For a given algorithm, given starting nodes of agents and a given 
sequence of adversarial decisions in an embedding of a graph $G$, a
rendezvous occurs,
if both agents are at the same point of the embedding at the same time.
Rendezvous is {\em feasible} in a given graph, if 
there exists an algorithm for agents such that for any embedding
of the graph, any (adversarial) choice of two distinct labels of agents, any starting nodes
and any sequences of adversarial decisions, the rendezvous does occur. The {\em
cost} of rendezvous is defined as
the worst-case number of edge traversals by both agents (the last partial
traversal counted as a complete one for both agents),
where the worst case is taken over all decisions of the adversary.

\subsubsection{Labeled agents}

The above asynchronous model was introduced in  \cite{DGKKPV} where the authors consider labeled agents. They study asynchronous rendezvous in the infinite line, in the
ring and in arbitrary connected graphs, both in the case when the initial instance $D$ between the agents is known, and when it is unknown. In the first two cases they propose several algorithms and analyze their cost. In one situation, for rendezvous in a ring of known size $n$ (but unknown $D$) they propose an algorithm of cost
$O(nl)$, where $l$ is the logarithm of the smaller label. This cost is optimal. The cost of their rendezvous algorithms in the infinite line has been subsequently improved in \cite{Sta09}. The rendezvous algorithms from  \cite{DGKKPV}  for the infinite line and for the ring are based on the following idea:
first transform the label $L$ of the agent in an appropriate way, and then {\em execute} the transformed label by making
some prescribed moves, if the current bit of it is 0 and making symmetric moves otherwise.

However, from the hindsight, the most influential part of  \cite{DGKKPV} was that concerning rendezvous in arbitrary graphs.
Here the authors tackle the question of feasibility. 
They prove that rendezvous is feasible, if an upper bound on
the size of the graph is known. As an open problem, the authors
state the question if asynchronous deterministic 
rendezvous is feasible in arbitrary graphs of unknown size.  The  solution from  \cite{DGKKPV}
heavily uses the knowledge of the upper bound on the size.

The general problem of feasibility of asynchronous  rendezvous for arbitrary graphs is solved in \cite{CLP}. The authors propose an
algorithm that  accomplishes asynchronous rendezvous in any connected countable (finite or infinite) graph, for arbitrary starting nodes.
A consequence of this result is a strong positive answer to the above mentioned open problem from  \cite{DGKKPV}: not only is rendezvous always possible, without the knowledge of any upper bound on the size of a finite (connected) graph, but it is also possible for all infinite (countable and connected) graphs.

The rendezvous algorithm from  \cite{CLP} is based on the notion of a {\em tunnel}.
Consider any graph $G$ and two routes $R_1$ and $R_2$ starting at nodes $v$ and $w$, respectively.
(These are sequences of edges, such that consecutive edges in the sequence are incident.)
We say that these routes form a tunnel, if there exists a prefix $[e_1,e_2,\dots,e_n]$ of route $R_1$ 
and a prefix $[e_n, e_{n-1},\dots, e_1]$
of route $R_2$, for some edges $e_i$ in the graph, such that $e_i=\{v_i,v_{i+1}\}$, where $v_1=v$ and $v_{n+1}=w$. 
Intuitively, the route $R_1$ has a prefix $P$ ending at $w$ and the
route $R_2$ has a prefix which is the reverse of  $P$, ending at $v$.
It is proved in \cite{CLP} that if routes $R_1$ and $R_2$ form a tunnel, then rendezvous is guaranteed, regardless of the decisions of the adversary.

A high-level idea of the algorithm from \cite{CLP} is to force the routes of any two agents with different labels to form a tunnel,
for every possible combination of starting nodes and (distinct) labels of the two agents. This is done by enumerating all quadruples 
 $(i,j,s',s'')$ , where $i$ and $j$ are different positive integers and $s'$, $s''$ are finite sequences of natural numbers, and arranging
 them in one countably infinite sequence. This enumeration is part of the algorithm and is the same for all agents. Then each agent processes quadruples 
 $(i,j,s',s'')$ in the order of their enumeration. Any starting configuration of agent with label $i$ placed at node $v$ and of agent with label
$j$ placed at node $w$ by the adversary corresponds to a quadruple $(i,j,s',s'')$, 
where $s'$ is a sequence of port numbers coding a path from $v$ to $w$ and $s''$ is a 
sequence of port numbers coding the reverse path from $w$ to $v$. During the processing of a quadruple  $(i,j,s',s'')$, a suffix is added to 
the already constructed initial segment of the route of agents with label $i$ or $j$, in such a way that if labels and initial positions of agents correspond to this quadruple, then the
routes of the agents form a tunnel. Since for some quadruple this condition must hold, arbitrary agents placed at arbitrary initial positions in the graph must
eventually meet.

The cost of the algorithm depends on the enumeration of the quadruples, and more
precisely on the position (in this enumeration) of
the quadruple  corresponding to the initial configuration of
the agents. During each phase of the algorithm, the length of the
routes of the two agents corresponding to the currently processed quadruple, is
at least doubled. Hence, the complexity of the algorithm is at least
exponential in terms of the number of quadruples with the same labels as those of the
two agents, that are before the quadruple corresponding to the initial configuration of
the agents in the enumeration. 
Thus the authors conclude their paper with the following natural question:

Does there exist a deterministic asynchronous rendezvous algorithm, working for all unknown connected finite graphs, with complexity polynomial in
the labels of the agents and in the size of the graph?

This question is answered in \cite{DPV} where the authors propose a deterministic asynchronous rendezvous algorithm, working for all unknown connected finite graphs, with cost polynomial in the size of the graph and {\em in the logarithm} of the smaller label.

The high-level idea of the algorithm from \cite{DPV}  is based on the following observation.
If one agent follows a trajectory traversing all edges of the graph during some time segment, then it must either meet the other agent or this other agent must perform at least one complete edge traversal during this time
segment, i.e., it must make {\em progress}.  
A naive use of this observation leads to the following simple algorithm (which is similar to that from \cite{DGKKPV}).
Let $R(n,v)$ be a trajectory starting at 
$v$ and traversing all edges of the graph of size at most $n$, and let $\overline{R(n,v)}$ be the reverse trajectory.
$R(n,v)$ can be , e.g., based on Reingold's \cite{R08} exploration.  An agent with label $L$ starting at node $v$ of a graph of size $n$ follows the trajectory $(R(n,v)\overline{R(n,v)})^{(2P(n)+1)^L}$, where $P(n)$ is an upper bound on the length of $R(n,v)$,
and stops.
Indeed, in this case the number of trajectories $R(n,v)\overline{R(n,v)}$ (that traverse all edges of the graph) performed by the larger agent (i.e., the agent with the larger label) is larger than the number of edge traversals by the smaller agent and consequently, if they have not met before,  the larger agent must meet the smaller one after the smaller agent stops, because the larger agent will still perform at least one entire trajectory afterwards.
However,  this simple algorithm has two major drawbacks.
First, it requires knowledge of $n$ (or of an upper bound on it) and second, it is exponential in $L$, while the goal is an algorithm {\em polylogarithmic} in $L$. Hence the above observation has to be used in a much more subtle way. As the authors of \cite{DPV} say, their algorithm ``constructs a trajectory for each agent, polynomial in the size of the graph and polylogarithmic in the shorter label, i.e., polynomial in its length, which has the following {\em synchronization} property that holds in a graph of arbitrary unknown size. 
When one of the agents has already followed some part of its trajectory, it has either met the other agent, or this other agent must have completed
some other related part of its trajectory. (In a way, if the meeting has not yet occurred, the other agent has been ``pushed'' to execute some part of its route.) The trajectories are designed in such a way
that, unless a meeting has already occurred, the agents are forced to follow in the same time interval such parts of their trajectories that meeting is inevitable. A design satisfying this
synchronization property is difficult due to the arbitrary behavior of the adversary and is the main technical challenge of the paper.''

The aim of \cite{DP3} is to investigate the difference of cost between the synchronous and asynchronous versions of a task executed by mobile agents.
The authors show that for some natural task executed by
mobile agents in a network, the optimal cost of its deterministic solution in the asynchronous setting has higher order of magnitude than that in the synchronous scenario. 
The task for which  this difference is proved is rendezvous of two agents in an infinite oriented grid. They consider two 
agents starting at a known distance $D$ in the infinite oriented grid. Agents do not have any global system of coordinates. They have to meet in a node or inside an edge of the grid, and the cost of a rendezvous algorithm is the number of edge traversals by both agents until the meeting. It is proved that in the synchronous scenario rendezvous
can be performed at cost $O(D\ell)$, where $\ell$ is the length of the binary representation of the smaller label, while cost $\Omega(D^2 +D\ell)$ is needed for asynchronous completion of rendezvous. Hence, for instances with $\ell=o(D)$, the optimal cost of asynchronous rendezvous is asymptotically larger than that of synchronous rendezvous.

\subsubsection{Anonymous agents}

The papers  \cite{BCGIL,CCGL} were among the first to consider asynchronous rendezvous of anonymous agents in graphs. The authors concentrate on particular graphs and use  strong additional  assumptions.
They consider rendezvous in an infinite two-dimensional grid, where ports are consistently labeled  $N,E,S,W$, and
agents know their initial coordinates in the grid, with respect to a common system of coordinates.
Hence, it is even problematic if such agents can be called anonymous (identical), as they are right away differentiated by their different initial coordinates.
It is proved in \cite{CCGL} that under these assumptions asynchronous rendezvous  can be accomplished at  cost $O(d^{2+\epsilon})$, where $d$ is the initial distance between the agents in the grid,
and $\epsilon$ is an arbitrary positive real. This result has been generalized and strengthened in \cite{BCGIL}, under the same assumptions. The authors show an asynchronous rendezvous
algorithm working for $\delta$-dimensional infinite grids with cost $O(d^{\delta} polylog(d))$. This complexity is close to optimal, as $\Omega(d^{\delta})$ is a lower bound on the
cost of any asynchronous rendezvous algorithm in this setting.

The problem of feasibility of asynchronous rendezvous  of anonymous agents in arbitrary graphs is solved in \cite{GP}. The authors show that rendezvous is possible if and only if 
 the views  from the initial positions of the agents are different, or these positions are connected by a path whose corresponding sequence of port numbers is a palindrome. The authors provide an algorithm guaranteeing deterministic asynchronous rendezvous from all such initial positions in an arbitrary connected graph that is either finite of arbitrary unknown size, or (countably) infinite.
 
 The algorithm is based on the idea of creating a tunnel, similarly as in \cite{CLP}. However, 
the main difficulty in designing the algorithm in the present setting is that, as opposed to  \cite{CLP},
agents do not have distinct labels allowing them to break symmetry. Hence symmetry can
be broken only by inspecting the views of the agents, if these views are different.
Even when they are different, the agents cannot know how deeply their views
have to be explored to find the first difference. Thus the algorithm proceeds in
epochs: in each consecutive epoch each agent explores its view more deeply, and
creates a code of this truncated view, subsequently treating it as its temporary
label and applying the procedure from \cite{CLP} to a restricted
list of quadruples. If views are different, a tunnel will be eventually created after
an epoch with sufficiently high index because in this epoch 
the truncated views serving as temporary labels of the agents will be different, 
and the argument from \cite{CLP} will work. The algorithm in \cite{GP}  has an additional feature
permitting the creation of a tunnel when views of the agents are the same but their
initial positions are joined by a path which is a palindrome. The simplest example of such a situation is 
the two-node graph with agents starting at extremities of the single edge. Views of the agents are the same and the code of the unique simple path
joining the initial positions of the agents is the palindrome $(00)$.
 
 \section{Approach in the plane}
 
 In this section we study the problem of approach in the plane or in terrains that are subsets of the plane: agents modeled as points moving in the terrain and starting at distinct points of it have to get at distance at most 1 of each other. Throughout the section we assume that agents have labels that are different positive integers. Each agent is equipped with a compass and a unit of length.
 
 The problem of approach in the plane of agents that have coherent compasses and the same unit of length can be reduced to the problem of rendezvous in an infinite oriented grid, where rendezvous is defined as getting at the same time to the same node or the same point of some edge of the grid. This means that if rendezvous in the infinite oriented grid can be solved under some set of assumptions about the agents, then the problem of approach in the empty plane can be solved under analogous
 assumptions.
 
 The problem of feasibility of asynchronous rendezvous in the plane or terrains is solved in \cite{CLP}. Consider any terrain (bounded or unbounded) that is a (topologically) closed subset of the plane with path-connected interior. (The latter means that for any two interior points of the terrain there exists a path, all of whose points are interior points of the terrain, connecting them). Agents start at arbitrary interior points of the terrain and their trajectories can be any polygonal lines. The authors propose an algorithm that accomplishes approach of any such agents in the terrain in finite time. Compasses and units of length of the agents may even be different.
 
 While the algorithm from  \cite{CLP} works in a very general setting, its drawback is the complexity: similarly as for the rendezvous algorithm for graphs proposed in this paper, the cost of the algorithm for terrains cannot be controlled. In order to find a more efficient algorithm, the authors of \cite{DP2} consider a scenario of restricted asynchrony. 
 Agents have coherent compasses and the same measure of length and of time, but they are assigned arbitrary, possibly different velocities by 
 an adversary. An agent can stay inert for a chosen amount of time, or it can move in a chosen direction and distance at its assigned speed. Under these restrictions the  authors propose an algorithm accomplishing approach in the plane in time polynomial in the unknown initial distance between the agents, in the logarithm of the smaller label and in the inverse of the larger speed. 
 The distance travelled by each agent until approach is polynomial in the first two parameters and does not depend on the third.
 
 The problem of tractable approach in the plane under full asynchrony has been finally solved in \cite{BBDDP}. The authors propose an algorithm accomplishing approach in the plane for agents whose speed may vary adversarially. The cost of the algorithm is polynomial in the initial distance between the agents and in the logarithm of the smaller label.
 
It should be mentioned that the result from \cite{BCGIL} concerning asynchronous rendezvous  in the infinite oriented two dimensional grid, accomplished at  cost $O(d^2 polylog(d))$, where $d$ is the initial distance between the agents in the grid, carries over to the task of approach in the plane under similar strong assumptions as in \cite{BCGIL}: agents have coherent compasses and the same unit of length and they know their initial coordinates in the plane, with respect to a common system of coordinates. Similarly as for the grid, this complexity is close to optimal, due to the  lower bound $\Omega(d^2)$.
 
 In \cite{EP2}, the authors consider the problem of approach in the plane under the synchronous model. 
 Agents are equipped with coherent compasses and the same unit of length, and have synchronized clocks. They make a series of moves. Each move specifies the direction and the duration of moving.
In a null move an agent stays inert for some time, or forever.
In a non-null move agents travel at the same constant speed, normalized to 1.

The twist of the model in this paper is restricted feedback that the agents get after each move, that is similar in spirit to the model of distance-aware agents from \cite{DDKU}, but weaker.
It is assumed that agents have sensors enabling them to estimate the distance from the other agent, but not the direction towards it. The authors consider two models of estimation.
In both models an agent reads its sensor at the moment of its appearance in the plane and then at the end of each move. This reading (together with the previous ones) determines the decision concerning the next move.
In both models the reading of the sensor tells the agent if the other agent is already present. Moreover,
in the {\em monotone model}, each agent can determine, for any two readings in moments $t_1$ and $t_2$, whether the distance from the other agent at time $t_1$
was smaller, equal or larger than at time $t_2$. It does not, however, get the value of this distance. In the weaker {\em binary model}, each agent can find out, at any reading, whether it is at distance less than $\rho$ or at distance at least $\rho$ from the other agent,
for some real $\rho>1$ unknown to them. To motivate their model, the authors mention that such distance estimation can be implemented, e.g., using chemical sensors. Each agent emits some chemical substance (scent), and the sensor
of the other agent detects it, i.e., the other agent {\em sniffs}. The intensity of the scent decreases with the distance. In the monotone model it is assumed that the sensors of the agents are very precise and 
can measure any change of intensity. In the binary model it is only assumed that the sensors can detect the scent below some distance (without being able to measure intensity or its changes) above which
the density of the chemical is too weak to be detected.

The authors investigate how the two ways of sensing influence the cost of meeting, defined as the total distance travelled by both agents until the meeting.
For the monotone model they present an algorithm achieving meeting in time $O(D)$, where $D$ is the initial distance between the agents.
This complexity is of course optimal. For the binary model they show that, if agents start at a distance smaller than $\rho$ (i.e., when they can sense each other initially) then
meeting can be guaranteed at cost $O(\rho\log \lambda)$, where $\lambda$ is the larger label, and that this cost cannot be improved in general.
It is also observed that,  if agents start at distance $\alpha\rho$, for some constant $\alpha >1$ in the binary model, then sniffing does not help, i.e., the worst-case optimal meeting cost is of the same
order of magnitude as without any sniffing ability.

In \cite{CKP3}, the authors consider both the task of approach and that of (exact) rendezvous of two agents in a terrain. 
Exact rendezvous (getting to the same point of the terrain at the same time) is possible because the  terrain is a polygon with polygonal holes,
and hence exact meeting can take place at the boundary of the terrain or of  a hole. Movements of the agents are asynchronous and agents have bounded memory: they are modeled as finite automata. The authors compare the feasibility of the task of rendezvous to that of approach for anonymous and for labeled agents. This gives rise to four scenarios, and
the authors show classes of polygonal terrains which distinguish all pairs of them from
the point of view of feasibility of rendezvous. The characteristics of the terrain that influence the feasibility of rendezvous and of approach include symmetries of the terrain, boundedness of its
diameter, and the total number of vertices of polygons in the terrain.

\section{Conclusion}

In this chapter we surveyed algorithmic results concerning deterministic rendezvous in networks and deterministic approach in terrains of the plane.
The appearance of many new results in the last few years is an indication of how vibrant is this domain of distributed computing. On the other hand, hopefully we managed to show that our understanding of rendezvous problems is still very incomplete, and a lot remains to be done. We would like to conclude the chapter by pointing out several avenues of research that  this author finds promising. This choice of the open problems is very subjective and reflects the personal taste of the author,
rather than their  importance on some hypothetical objective scale.

It seems reasonable to classify possible open problems into two categories. The first category is strengthening of the existing results without changing the model
under which they were originally obtained. Here the most interesting problems seem those aiming at improving the efficiency of existing algorithms and ultimately 
finding an algorithm of optimal complexity. In this category we would put forward the problem of finding: 
\begin{itemize}
\item
an optimal-time synchronous rendezvous algorithm in arbitrary graphs
\item
an optimal-cost asynchronous rendezvous algorithm in arbitrary graphs
\item
an optimal-cost asynchronous approach algorithm in the plane
\end{itemize}

All these problems are formulated for labeled agents. The third problem is for agents with coherent compasses and the same unit of length.
As we know, for all these problems polynomial algorithms are known, but, especially in the case of problems 2 and 3, the exponents of the polynomials are very large. Finding algorithms of optimal complexity for any of these scenarios seems to be very challenging. We think that even a significant improvement of the existing algorithms would be a big step forward.

The second category concerns investigating rendezvous under new models. It seems that the interplay between efficiency of rendezvous or approach and the communication capabilities of agents is still poorly understood. This problem has been ``touched'' in papers \cite{DDKU,EP,EP2} but a more complete analysis
of tradeoffs between communication of agents and efficiency of rendezvous under various scenarios is badly needed. A realistic type of communication, especially
for agents in the terrains, seems to be wireless. This can be challenging, especially for large teams of agents, as usual problems of wireless communication concerning collisions would have to be tackled.

Another interesting issue are trade-offs between memory of the agents and their sensory capabilities.
In this chapter we assumed that agents cannot see other agents prior to meeting but usually they have significant or even unbounded memory. 
By contrast, in chapter XXX, it is usually assumed that agents cannot remember any information
from the previous Look-Compute-Move cycles, but they can take a snapshot of the entire network (or a large part of it) during the Look action. 
Most applications are probably in between those two extremes: agents have some memory of past events (for example of constant size), but their sensory capabilities are more limited,
e.g, they can only perceive the part of the configuration at a given radius from their current position, they cannot see ``through'' other agents, etc.  
Studying feasibility of rendezvous under such more balanced
scenarios could involve characterizing initial configurations for which rendezvous (or approach in the plane) is possible, and trying to optimize the cost of meeting, which,
under very small memory of the agents, is often still unknown.



\end{document}